\documentclass[aps,pre]{revtex4}

\newlength{\defbaselineskip}
\def\ci{\hfill \vline &}
\def\Ang{$\mathrm{\AA}\ $}

\begin{document}

\title{Approximating nonequilibrium processes using a
collection of surrogate diffusion models}

\author{Christopher P. Calderon}
\affiliation{
Department of Statistics and Department of Computational and
Applied Mathematics, Rice University, Houston, TX 77005-1892, USA.
}

\author{Riccardo Chelli}
\affiliation{
Dipartimento di Chimica, Universit\`{a} di Firenze, Via della
Lastruccia 3, I-50019 Sesto Fiorentino, Italy, and European
Laboratory for Non-linear Spectroscopy (LENS), Via Nello Carrara
1, I-50019 Sesto Fiorentino, Italy. }

\date{\today}

\begin{abstract}
The surrogate process approximation (SPA) is applied to model the
nonequilibrium dynamics of a reaction coordinate (RC) associated with
the unfolding and refolding processes of a deca-alanine peptide at 300
K. The RC dynamics, which correspond to the evolution of the
end-to-end distance of the polypeptide, are produced by steered
molecular dynamics (SMD) simulations and approximated using overdamped
diffusion models. We show that the collection of (estimated) SPA
models contain structural information ``orthogonal'' to the RC
monitored in this study. Functional data analysis ideas are used to
correlate functions associated with the fitted SPA models with the
work done on the system in SMD simulations. It is demonstrated that
the shape of the nonequilibrium work distributions for the unfolding
and refolding processes of deca-alanine can be predicted with
functional data analysis ideas using a relatively small number of
simulated SMD paths for calibrating the SPA diffusion models.
\end{abstract}


\maketitle

\section{Introduction}

Single-molecule experiments \cite{physicstoday} offer the possibility
of exploring the dynamical responses in systems without having to
resort to ensemble measurements
\cite{jarzynskiCROOKSVER,kiang07,voth07,schultenJCP04,procacci}. In
recent years these types of investigations have received increasing
attention, due partially to developments in nonequilibrium statistical
mechanics \cite{jarzynski97,crooks98}.  In the context of protein
folding, vastly different dynamical trajectories can be observed when
monitoring the nonequilibrium time evolution of a reaction coordinate
(RC) obtained from independent single-molecule measurements carried
out at identical experimental conditions. The differences in dynamical
response cannot usually be attributed only to classical thermal noise,
but also to the different atomistic ``collective'' coordinates
\cite{willy06} (e.g., radius of gyration) consistent with the
conditions of the system at the initial time of the various
nonequilibrium experiments. In many cases, these types of collective
variables strongly influence the nature of observed dynamical
response. This type of phenomenon has been quantified by modeling with
``multiple-states" \cite{zhuang02, fernandezPNAS07,wangLigand_06} or
using the so-called ``dynamic disorder" description \cite{xie07}.
Often, both descriptions can reasonably be thought of as artifacts
resulting from an imperfect RC
\cite{wangLigand_06,Dobson_science06AdK}.

One concrete example that exhibits the concepts expressed above is the
extensively studied \cite{schultenJCP04,procacci} process of unfolding
and refolding of deca-alanine (a deca-peptide consisting of alanine
residues). In Fig. \ref{fig:rcfail} we report some snapshots of
deca-alanine taken from two forced refolding realizations along the
end-to-end distance (the RC) produced by steered molecular dynamics
(SMD) simulations \cite{procacci}. By comparing the time evolution of
the RC in the two computational experiments (see panels in the
figure), one would be inclined to classify the two nonequilibrium
trajectories as (statistically) identical, or at least very
similar. However, when the snapshots of the entire molecule are
observed, it is easy to distinguish between the dynamical responses
(``proper folding" and ``misfolding"). Another instance of this type
of phenomenon is documented in Ref. \onlinecite{ioanFR07}. This
situation is relevant to a variety of different systems because, when
all-atom descriptions are used, it is usually difficult to select RCs
which accurately reflect all of the relevant details of the system
\cite{hummerPNAS01,schultenJCP04,grubmuller}.  Even if good RCs are
known and easily manipulatable, \emph{a priori} quantitative knowledge
of how the system responds to external stimuli that vary these RCs is
usually lacking.

This paper aims at providing this type of information by applying
pathwise time series techniques \cite{raoDIFF,SPA1} to the output of
nonequilibrium single-molecule computer simulations. The numerical
procedures presented here also have applicability to wet-lab times
series data \cite{kiang07,fernandezREVIEW00,jarzynskiCROOKSVER}.
Nonequilibrium simulations are attractive because they sometimes allow
a researcher to obtain detailed equilibrium information about portions
of phase space that, if the system were unforced, would not be
explored in the typical simulation timescale
\cite{hummerPNAS01,schultenJCP04,kosztin06,grubmuller}.  The system
studied throughout is a SMD simulation of deca-alanine in vacuo
maintained at constant temperature by a Nos\'{e}-Hoover thermostat
\cite{hoover85}. We systematically manipulate the end-to-end distance
RC of the deca-alanine peptide, referring to it as the ``steered" RC
\cite{schultenJCP04}.

The purpose of this study is two-fold. The first goal is to present a
computational approach which attempts to learn about conformational
dynamics not directly contained in a single trajectory of the steered
RC (but is indirectly contained in a collection of
trajectories). Several configurational degrees of freedom, such as
dihedral angles, may be indeed important in determining the dynamics
of the studied system. These degrees of freedom can be thought of as
``other'' RCs and are not directly included in the dynamical models
calibrated from the observed data. However, these RCs can be
relatively slowly evolving (compared to the timescale of the
simulation) and may therefore effectively modulate the dynamics of the
steered RC.  Because of these other RCs, independent trajectories may
be characterized by significantly different dynamics yielding a
\emph{collection} of diffusion models \footnote{Or more generally, a
collection of dynamical responses (one is free to use dynamical models
more complicated than diffusion processes).} . In this study, ideas
from functional data analysis (FDA) \cite{ramsay} are used in our
models to include this important source of ``dynamical
heterogeneity''. In particular we develop a procedure, that we call
``FDA bootstrapping'' (which is a specific instance of ``type II
bootstrapping'' discussed in Ref.  \onlinecite{SPA1}).  This
bootstrapping is able to correlate the work done on the system during
an unfolding or refolding realization to appropriate functions
associated with a surrogate process approximation (SPA) model
\cite{SPA1}. This type of treatment is attractive because it does not
require the identification or manipulation of the slowly evolving
collective coordinates (this is particularly relevant to wet-lab
studies).

The second aim of the paper is to show how the proposed approach
(coupled with ideas from Ref. \onlinecite{SPA1}) can be applied to
recover the non-Gaussian work distributions associated with
nonequilibrium SMD folding/unfolding of deca-alanine. This may
possibly assist experiments or simulations where path samples are so
expensive or time consuming that obtaining an adequate statistical
sample is difficult.  The statistical validity of approximating
deterministic (chaotic) dynamics using diffusion approximations has
also been addressed here given its relevancy to a variety of atomistic
systems \cite{evans07,hoover85,kushner}. However, in order to keep the
presentation streamlined, we have relegated these results to the
Supplementary Material \footnote{See EPAPS Document No. ???? for all
references to Supplementary Material. This document can be reached
through a direct link in the online article's HTML reference section
or via the EPAPS homepage (http://www.aip.org/pubservs/epaps.html).}.

The outline of the article follows.  In Section \ref{sec:methods} we
review the basic concepts underlying the SPA model and the FDA
methodology and provide the computational details regarding the SMD
simulations of deca-alanine.  In Section \ref{sec:results} we report
the results regarding the determination of the work distributions
using diffusion models. In this section we also report data and
discuss issues regarding the connection between the dynamical
information from SPA models and other possible RCs ``orthogonal"
\footnote{We simply define two reaction coordinates as being
``orthogonal" when the information set is disjoint (i.e. the value of
each coordinate contains relevant system information that cannot be
determined from the value of the other coordinate) } to the steered
one. A summary and some potential future directions of research are
finally discussed in Section \ref{sec:conclusions}.

\section{Theory and methods}
\label{sec:methods}

\subsection{SMD simulations of deca-alanine}
\label{sec:compdetails}

As stated in the Introduction, the dynamical system considered in the
present study is the process of unfolding (and refolding) of one
deca-alanine peptide at finite temperature. As in a previous work
\cite{procacci}, SMD simulations were used as a device for producing
the forced nonequilibrium dynamics of the steered RC, $Z$, which is
taken to be the end-to-end distance of deca-alanine.  More
specifically, $Z$ corresponds to the distance between the N atom of
the N-terminus amino-acid (constrained to a fixed position) and the N
atom of the C-terminus amino-acid (constrained to move along a fixed
direction). The values of $Z$ in the completely folded and unfolded
states of deca-alanine are assumed to be 15.5 and 31.5 \Ang,
respectively. In the former state, deca-alanine is found in a
$\alpha$-helix form, while an elongated ``random coil"
\cite{schultenJCP04} configuration characterizes the latter state. We
have arbitrarily assumed the unfolding and refolding processes as
forward ($F$) and reverse ($R$), respectively.

It should be noted that in general the end-to-end distance does not
determine uniquely the configurational state of polypeptides.
However, in the specific case of deca-alanine, the equilibrium
distribution at $Z = 15.5$ \Ang corresponds to an ensemble of
microstates tightly peaked around the $\alpha$-helix form, as for this
end-to-end distance alternative structures are highly unlikely under
the associated stationary distribution \footnote{We refrain from using
the term ``equilibrium distribution " because a biasing potential is
used to maintain the configurations at the desired $Z$ value.}.  The
same holds true for the state corresponding to $Z = 31.5$ \Ang
(i.e. there is a tightly peaked stationary probability density
corresponding to an ensemble of enlongated random coils
configurations).  Because of this, the thermodynamically relevant
phase space points consistent with the $\alpha$-helix (and elongated
random coil) to be used as initial microstates of the SMD simulations
can readily be obtained by sampling from the stationary distribution
associated with a constrained RC molecular dynamics
simulation. However, as discussed in the Introduction, subtle
configurational differences in these initial microstates may yield
vastly different dynamical behaviors when the system is driven out of
equilibrium.  The initial microstates of deca-alanine for the $F$ and
$R$ realizations were randomly drawn from two ``equilibrated" biased
molecular dynamics simulations constraining $Z$ to 15.5 and 31.5 \Ang,
respectively, by means of a stiff harmonic potential (force constant
$k_\mathrm{pull}$ equal to 800 kcal mol$^{-1}$ \AA$^{-2}$). In both
equilibrium simulations and in the subsequent SMD simulations,
constant temperature (300 K) has been enforced using a Nos\'e-Hoover
thermostat\cite{hoover85}. Force field parameters have been taken from
Ref. \onlinecite{mackerell98}. For each type of process, $F$ and $R$,
we have generated 10$^3$ realizations guiding $Z$ from 15.5 to 31.5
\Ang ($F$ paths) or from 31.5 to 15.5 \Ang ($R$ paths) using a
time-dependent harmonic potential with the force constant reported
above. The parameter driving the reaction coordinate, that we denote
with $\lambda(t)$, evolves in time with an externally determined
constant rate of 0.032 \AA~ ps$^{-1}$ (the same rate is used for both
$F$ and $R$ processes). The harmonic guiding potential $V(t)$ is the
following: $V(t) = 0.5 ~ k_\mathrm{pull} ~ [ Z(t) - \lambda(t)
]^2$. We stress that $\lambda(t)$ is a deterministic function common
to all SMD simulations and $Z(t)$ represents a stochastic
process. Moreover, we note that, in order to generate the large amount
of temporal data needed to reliably calibrate diffusion models and
still have the time spacing between observations large relative to the
faster timescales
\cite{maria06,zwanzig,stuartHYDRO,stuart06,llglassy}, we have
performed longer SMD simulations than those reported in
Ref. \onlinecite{procacci}.

The ORAC program was used in all simulations\cite{procacci97}. It uses
a multiple timestep technique, r-RESPA, to integrate the equations of
motion.  In particular forces between atoms whose distance is greater
than 12 \Ang have been updated every 5 fs; forces between atoms whose
distance falls in the range 6-12 \Ang have been updated every $\sim$
1.667 fs; forces between atoms distant less than 6 \Ang, torsional
forces, and forces between atoms in 1-4 position have been updated
every $\sim$ 0.556 fs; bending and stretching forces have been updated
every $\sim$ 0.278 fs. Simulation data needed for modeling, i.e.,
external work and $Z(t)$, have been stored every 50 fs.

\subsection{Surrogate Process Approximation and  Bootstrapping}
\label{sec:modboot}

If we assume that only a relatively small number of paths or
realizations are available from an experiment/calculation, it is
natural to try to use the rich amount of dynamical data contained in
the paths in order to calibrate a surrogate model and then turn to
this model to augment the limited data set for improving the quality
of the sample. This is the basic idea behind the SPA method used in
the present work and introduced in Ref. \onlinecite{SPA1}. We attempt
to fit time series data coming from constant velocity SMD simulations
of deca-alanine to time dependent overdamped diffusion models using
locally parametric maximum likelihood methods.  The underlying complex
system used to generate the time series is a system of deterministic
chaotic ODEs, as opposed to a Langevin type process which was
originally used in Ref. \onlinecite{SPA1}.

Specifically, a single RC trajectory obtained from a SMD simulation is
used to calibrate a collection of local diffusion models, each model
being characterized by a diffusion and a drift coefficient
\cite{SPA1}. For the transition density expansions, we utilize the
Simulated Maximum Likelihood \cite{sml} estimator.  Local overdamped
diffusion models are estimated at the ``center points"
$\{15,16,\ldots,32\}$ \Ang of the RC (for a total of 18 local models)
using a neighborhood radius of 0.6 \Ang for each local model
\cite{SPA1}. The diffusion and the drift coefficients of the local
diffusion models are then stitched together to form two continuos
functions, i.e., the diffusion and the drift functions representing
the global diffusion model of the RC trajectory. These functions are
obtained using a piecewise polynomial smoothing routine available from
Matlab's spline toolbox ($\mathrm{csaps}$, with smoothing parameter
equal to 0.4). In Fig. \ref{fig:multmodel} we show the drift and
diffusion functions calibrated using 15 forward and reverse SMD
trajectories of deca-alanine.  It should be noted that the diffusion
model fit to the data is identical to that used in
Ref. \onlinecite{SPA1}. Section II of the aforementioned reference
provides a more comprehensive discussion on the diffusion models.

The resulting global diffusion model can then be used to generate a
collection of new paths taking the same initial condition as the SMD
trajectories, but using a batch of random number sequences.  The
different random number sequences used to create Brownian paths are
intended to incorporate variability that can be attributed to
physically unimportant fast degrees of freedom, such as vibrational
degrees of freedom associated with covalent bonds, present in the full
atomistic system. The term ``type I bootstrapping" is used here to
describe the generation of new paths from the procedure described
above.

There is interest in understanding the consequences of approximating
highly structured chaotic ``noise" with a fairly crude diffusion model
\cite{stuart06,stuartHYDRO,llglassy} when the data are sampled
frequently in time (see the Supplementary Material for an extensive
discussion along these lines). In principle, any stochastic process
could be fit to the time series data and serve as the surrogate, e.g.,
underdamped diffusions \cite{zwanzig}, non-Markovian processes
\cite{maria06,xie04}, etc.  However, the motivation for focusing on
estimation of overdamped models stems from two facts. (1) In
macromolecules large collective motions typically \cite{willy06} cause
an effectively overdamped system even when the traditional
hydrodynamic ``damping" (or viscosity) is very low
\cite{ansari_odamp_99}.  (2) Typically, in wet-lab experiments, only
simple position-like coordinates are available, which force the
researcher into a framework where overly simplified dynamical models
are used because the information to calibrate more sophisticated
models is simply not accessible (or perhaps deemed not worth pursuing)
\footnote{Besides the statistical validity of the overdamped
diffusion approximation can be readily tested  using information
contained in our models (see Supplementary Material).}.

Characterizing the errors introduced by ignoring degrees of freedom in
a low-dimensional RC calculated from observations of a detailed
atomistic system is a difficult and open problem
\cite{xie04,maria06,stuartHYDRO}. The multiscale nature of the
signals that come from atomistic systems makes the legitimacy of using
a single low dimensional diffusion model questionable.  The approach
described in the next section aims at remedying this problem by
enhancing the sampling of \emph{models} that a batch of fully
atomistic paths sometimes yields.  A physics-based description of the
presence of a collection of models would assume that the effective
free energy \cite{stuart06} is a function of two or more RCs, but we
only observe and model the dynamics of one.  Unlike the RC explicitly
considered in the model, the ``other" RCs are not controlled either at
the beginning of the nonequilibrium experiment or during it. This
implies that each RC path may give rise to a dynamical model which
depends heavily on unobserved RCs. Therefore, many RC paths may result
in a collection of models, whose ``heterogeneous dynamical response"
can be attributed to the unconstrained other RCs in the nonequilibrium
experiment. In this paper, we present a technique for using the
information contained in this distribution of models to make refined
predictions of the work distribution from a small data set.  We also
demonstrate that this distribution of models can provide indirect
information about other important RCs. This is made possible by FDA
ideas.

\subsection{Using FDA Ideas to Bootstrap Work Paths}
\label{sec:FDA}

The basic goal of FDA \cite{ramsay,ferraty06} is to characterize the
distribution of a family of curves. Instead of statistically analyzing
finite dimensional random vectors (e.g., Euclidean vectors), one
focuses instead on the distribution of objects living in an infinite
dimensional space. In the present case we deal with the space of
continuous functions that describe the dynamics of the various SMD
trajectories. One may associate three functions to each SMD
trajectory: the work function corresponding to the time sequence of
the external work performed on the system during the simulation and
the two SPA coefficient functions (see Fig. \ref{fig:multmodel}). Here
we assume that the observed work functions are independent and
identically distributed (iid).

Unfortunately one usually does not have the luxury of continuous time
observation in many applications \cite{ramsay,raoDIFF}.  However, when
the underlying objects are believed to be relatively smooth functions
(as it seems to be the case of the deca-alanine system) it makes sense
to discretely sample the external work at a moderate sampling
frequency. Then one can attempt dimension reduction by using principal
components analysis (PCA) tools \cite{ramsay}.  To determine the
correlation between the SPA model functions and the observed work
functions, one could create a raw observation matrix of the form:

\begin{center} $  \left[ \begin{array}{ccccccc}
 \mathcal{W}^{(1)}({ \tau_1}) &  \mathcal{W}^{(2)}({\tau_1}) & \cdots &  \mathcal{W}^{(N)}({\tau_1}) \\
 \mathcal{W}^{(1)}({\tau_2}) &  \mathcal{W}^{(2)}({\tau_2}) & \cdots &  \mathcal{W}^{(N)}({\tau_2}) \\
&  \vdots & \\
  \mathcal{W}^{(1)}(\tau_{D}) &   \mathcal{W}^{(2)}(\tau_{D}) & \cdots &   \mathcal{W}^{(N)}(\tau_{D}) \\
  \vspace{9pt} \\
 \tilde{A}^{(1)}(Z(\tau_1)) &  \tilde{A}^{(2)}(Z(\tau_1)) & \cdots &    \tilde{A}^{(N)}(Z(\tau_1)) \\
  \tilde{A}^{(1)}(Z(\tau_2)) &  \tilde{A}^{(2)}(Z(\tau_2)) & \cdots &  \tilde{A}^{(N)}(Z(\tau_2)) \\
&  \vdots & \\
  \tilde{A}^{(1)}(Z(\tau_D)) &   \tilde{A}^{(2)}(Z(\tau_D)) & \cdots &  \tilde{A}^{(N)}(Z(\tau_D)) \\
\vspace{9pt} \\
 \tilde{C}^{(1)}(Z(\tau_1)) &  \tilde{C}^{(2)}(Z(\tau_1)) & \cdots &  \tilde{C}^{(N)}(Z(\tau_1)) \\
 \tilde{C}^{(1)}(Z(\tau_2)) &  \tilde{C}^{(2)}(Z(\tau_2)) & \cdots &  \tilde{C}^{(N)}(Z(\tau_2)) \\
&  \vdots & \\
  \tilde{C}^{(1)}(Z(\tau_D)) &   \tilde{C}^{(2)}(Z(\tau_D)) & \cdots &   \tilde{C}^{(N)}(Z(\tau_D)) \\

\end{array} \right]\equiv \left[ \begin{array}{c}
\mathbf{\mathcal{W}} \\ \\ \mathbf{ \tilde{A}} \\ \\
\mathbf{\tilde{C}} \end{array}\right], $
\end{center}

where $\mathcal{W}^{(i)}({\tau_j})$, $\tilde{A}^{(i)}(Z(\tau_j))$ and
$\tilde{C}^{(i)}(Z(\tau_j))$ are the work, drift and diffusion
functions at time $\tau_j$ corresponding to the $i$th SMD
trajectory. The $N$ matrix columns correspond to the function
``samples" observed discretely at the times $\tau_1$, $\tau_2$,
$\dots$ , $\tau_D$\footnote{It should be noted that the actual
reaction coordinate value ($Z(\tau)$) is very close to the
deterministic target value ($\lambda(\tau)$) because of the high
spring constant used in the SMD simulations.  In the FDA analysis
employed, there was a negligible difference between results obtained
using $Z$ and those obtained using $\lambda$.}. After projecting {
these functions} onto a basis, the weights are then ``stacked" upon
each other (this simply means that the weights of the different
projections are concatenated) to form a new matrix with blocks
$\mathbf{\mathcal{W}^{\prime}} , \mathbf{ \tilde{A}}^{\prime} $ and $
\mathbf{ \tilde{C}}^{\prime}$
\footnote{The matrix we work with in this study is actually a
block of the weights numerically obtained when our discretely
observed functions are projected onto a Haar wavelet basis set.}.
The specific  weight that should be assigned to the different
blocks is not easy question to answer because  introducing an
appropriate metric (or semimetric) is difficult \cite{ferraty06}
without \emph{a priori} information on the smoothness of the
curves. In this work we use a  fairly simple weighting procedure
made of the following steps: (1) within each block, the average
population (across the $N$ samples) of the coefficients (a vector)
was subtracted from the individual coefficient vectors; (2) within
each block, the coefficients resulting from step 1 were
``normalized" by the corresponding block standard deviation
\footnote{Within each block, the variance at a grid point was
first computed across the function population. The variance at all
grid points within the block was then summed and the square root
of this quantity is what we call the ``standard deviation".}. The
intention is to give each block the same relative importance in a
PCA type decomposition. After the PCA was carried out on  the data
matrix, each stacked observation (of basis weights) was then
projected onto  the constructed principal components basis (three
PCA modes were retained in all cases reported) and the PCA weights
of each projection  were then recorded. In order to get the
distribution of the weights two methods were used:

\begin{enumerate}
\item Under the ad hoc assumption that the PCA weights are
statistically independent \footnote{ Since the initial configurations
are iid samples from a stationary distribution, there is no
statistical correlation between the work, diffusion and drift
functions for two different SMD simulations.
\label{indcurve}} (by construction of the PCA decomposition, the
weights are linearly uncorrelated), the empirical distributions
(across the functional curve population) were recorded for each PCA
projection and new samples were created by resampling the PCA weight
distributions. This new set of weights augments the original small set
of weights (the basic idea being to ``interpolate" between function
curves by resampling from the empirical cumulative distribution
functions).
\item In an effort to computationally remove any dependence between
the weights, and still under the assumption given in Footnote 49, we
employed tools such as independent components analysis (ICA)
\cite{stone_stICA,fastICA} or projection-pursuit methods to transform
the weights into roughly ``decoupled" coordinates (we used the FastICA
package \cite{fastICA}).  We then empirically measured the cumulative
distribution function of the weights and used the resulting
distributions to generate a new set of weights. The resulting weights
were finally transformed back to the original coordinates and the
resulting work paths were recorded \footnote{Unfortunately ICA
typically requires very large samples sizes to decouple and separate
signals into (roughly) independent components.  However if one has a
large collection of paths and is interested in using ICA to classify
trajectory clusters, then ICA might be more useful (and reliable).}.
\end{enumerate}

The type of methods described above are referred to as ``FDA
bootstrapping". The basic steps of the overall method using one of the
above methods are summarized below:

\begin{itemize}
\item Let $\mathbf{X}\equiv [ \mathbf{\mathcal{W}^{\prime}}, \mathbf{
\tilde{A}^{\prime}}, \mathbf{\tilde{C}^{\prime}} ]^T$ and denote the
corresponding linear basis by $\{\mathbf{u_1},\mathbf{u_2},
\ldots\}$. Each column of $\mathbf{X}$ is then projected onto each
column of the linear basis to obtain an ensemble of weights $\equiv \{
\begin{array}{c} \mathbf{w_1}, \mathbf{w_2}, \ldots \end{array}\}$.
\item The empirical cumulative distribution function associated
with each (raw or processed) row of the $\mathbf{w_i}$ vectors is
determined (i.e. across the $N$ samples, a cumulative distribution
function of each component of the weight vector is measured).
\item Using the above cumulative distribution function, a large
number of ``bootstrapped" weights are then resampled.
\item Synthetic work paths and diffusion models are finally
produced using ``FDA bootstrapped models''. The cost of this
operation is usually marginal in relation to the overall data
production procedure.
\end{itemize}

\section{Results}
\label{sec:results}

\subsection{Failure of Direct Work Process Approximation}
\label{sec:workapproximaion}

SPA models were calibrated by observing 15 randomly selected SMD
paths. The time dependent distribution of $Z$ was accurately modeled
with our stochastic dynamical models (see Figs. 1-2 in the
Supplementary Material).  The accuracy in the $Z$ distribution
approximation gives us partial hope for using the estimated SPA models
to generate a collection of surrogate processes from which work can
\emph{directly} be measured by type I bootstrapping (as was done in
Ref. \onlinecite{SPA1}). 10$^4$ such paths were generated and the work
done on the system (pulling in the $R$ direction) was recorded as a
function of time.  The ``pulling" parameters used in these type I
bootstrapping SPA simulations were identical to those used in the
corresponding SMD paths.  For comparison, the work done on the system
as a function of time from the 1000 SMD paths has also been
computed. Both SPA and SMD average work profiles, along with their
standard deviations, are shown in Fig. \ref{fig:worktraj}. The
standard deviation of the SPA work paths is far too large compared to
the SMD process it aims at approximating.  This can be very
problematic if one wants to use the approximate work distribution to
compute free energy differences using a nonequilibrium method like the
Jarzynski equality\cite{jarzynski97}.

A few representative work paths are also plotted for both the SPA and
SMD cases. When individual paths are compared, one clearly observes
that the SPA work paths are too ``rough" in relation to the relatively
smooth SMD work paths. This smoothness is likely caused by structured
fast degrees of freedom not explicitly included in our surrogate
models.  Accurately approximating this structure would be a formidable
time series modeling problem and an interesting future research
direction.  However here we turn instead to the tools of FDA to model
the work process. The estimated SPA models are still useful (despite
the fact that they cannot directly simulate useful work paths) because
we demonstrate that the information in the collection of SPA models
indirectly summarizes some features of the complex SMD process.

\subsection{Applying FDA Bootstrapping to Approximate Work Distributions}
\label{sec:PMFest}

We recall that one of the motivations of this study is to use the
information contained in the estimated SPA models for approximating
the shape of the work distributions associated with SMD simulations.
In several early computational
studies\cite{ritortgausswork_03,schultenJCP04,kosztin06}, Gaussian
work distributions were observed in pulling nonequilibrium processes.
Most of the arguments in the literature that attempt to justify this
fact are either heuristic \cite{ritortgausswork_03} or make the
assumption that a \emph{single} overdamped diffusion can be used to
model the time series coming from nonequilibrium SMD simulation data
\cite{schultenJCP04}.  However it has been also shown that, even
seemingly Gaussian work distributions, may not be Gaussian. This is
just the case of the work distribution for the unfolding process of
deca-alanine whose shape, though very similar to a Gaussian profile,
is inherently asymmetric\cite{procacci}.  This brings up a subtle
point we want to stress. In this system a \emph{single} overdamped
model (calibrated at this timescale) is not adequate to describe the
dynamics of each nonequilibrium response. This is likely caused by
slowly evolving degrees of freedom outside of the selected reaction
coordinate that have not been sufficiently ``averaged out"
\cite{zwanzig}
\footnote{We do stress that the SPA models do accurately capture many
statistical features of the true data (see Supplementary
Material).}. In our system, this causes one to estimate
\emph{multiple} overdamped diffusion models.  The approximation we
introduced takes this model variability into account and should be
viewed as a new type of stochastic process that is different than a
single ``simple" overdamped diffusion. This new stochastic process
violates the conditions that guarantee a Gaussian work distribution
stated in Ref. \onlinecite{schultenJCP04}.  Because guaranteeing (or
enforcing) a Gaussian work distribution can be difficult with certain
reaction coordinates, it would be useful to have a tool which can be
used to check the validity of the Gaussian work assumption without
requiring a large number of nonequilibrium realizations.  In the
deca-alanine system, the FDA motivated techniques presented here are
shown to be useful in this context (see below).

The work distributions recovered from the 1000 SMD simulations and
from the various FDA bootstrapping approaches (see Section
\ref{sec:FDA}) for the $R$ realizations are shown in Fig.
\ref{fig:Rdens}. 15 SMD trajectories, taken randomly from the overall
ensemble of SMD trajectories, were used to calibrate 15 SPA models
used for the various FDA bootstrapping approaches. The work values
corresponding to these SMD trajectories are also reported in
Fig. \ref{fig:Rdens}. The FDA bootstrapping types are classified as
follows: (1a) the FDA bootstrapping used diffusion and drift
coefficient functions from the SPA models along with the corresponding
work function from SMD trajectories ($\mathcal{WAC}$ bootstrapping);
(1b) same as previous, but ICA decoupling has been used on weights
($\mathcal{WAC}$ ICA bootstrapping); (2a) only the work function from
the SMD trajectories has been used in the FDA bootstrapping
($\mathcal{W}$ bootstrapping); (2b) same as previous, but ICA
decoupling has been used on weights ($\mathcal{W}$ ICA bootstrapping).
Three PCA modes were used in all cases reported.  The number of
surrogate paths extracted from each FDA bootstrapping type is $10^4$.
In addition the same random number stream was used for each case in
order to minimize sampling errors and attribute differences in the
results solely to the systematic differences in the computational
methods.

We note that a naive bootstrapping ($\mathcal{W}$) falsely predicts
multiple modes, whereas the other bootstrapping schemes fairly
accurately capture the shape of the SMD work distribution.  This is
indeed quite surprising given that only 15 SMD paths underly the SPA
models. The fact that a structured tube of functions is observed in
the SPA models (sse Fig.  \ref{fig:multmodel}), and that using this
information in the FDA bootstrapping improves the reproduction of the
work distribution, suggests that the work and the diffusion dynamics
of the RC (accounted for by the SPA models) are indeed significantly
correlated.

Increasing the number of SMD paths underlying the SPA models improves
the estimation of the low order moments (see Table \ref{tab:Wstats}),
but the overall shape of the predicted work distributions does not
change substantially. We point out that $\mathcal{W}$ ICA and
$\mathcal{WAC}$ ICA methods seem to work better than $\mathcal{W}$ and
$\mathcal{WAC}$ methods because there are multiple clusters (discussed
in detail in Section \ref{sec:storytime}) of observed dynamical
responses ( i.e., SPA models) that a linear method like PCA cannot
directly account for \cite{fastICA,stone_stICA}. The SPA model
information aids in detecting these clusters , thus allowing for an
appropriate assignment of the weights to the work paths. Though the
weighting induced by the SPA functions resulting in an improved work
distribution estimation is just a fortuitous coincidence, in the
future, we hope to develop schemes which systematically correlate
these SPA ``function clusters" with slowly evolving collective degrees
of freedom and assign appropriate weights to the paths using
empirically determined correlations.

In the $\mathcal{WAC}$ and $\mathcal{W}$ bootstrapping methods, small
batches of 15 SMD paths were drawn (without replacement) and used to
calibrate 15 SPA models and then these models are used to approximate
the nonequilibrium work distribution. When this procedure is repeated
(with new batches of iid SMD sample paths) the mode variability in the
resulting work distribution is significant (see Table
\ref{tab:Wstats}).  The mean and mode in the $\mathcal{WAC}$ ICA
approach is relatively stable and when independent batches of 15 SMD
paths are used in FDA bootstrapping, the resulting work distributions
have non-Gaussian shape with roughly the same mode. This is
demonstrated in Fig.  \ref{fig:CEver}, where $\mathcal{WAC}$ ICA
bootstrapping is applied for calculating the $R$ and $F$ work
distributions using independent batches of SMD paths (3 batches of 15
paths for the $R$ direction and 2 batches of 15 paths for the $F$
direction).  For comparison, in the inset of Fig. \ref{fig:CEver} we
report the $F$ and $R$ work distributions recovered from the 1000 SMD
paths.  The free energy difference between folded and unfolded states
of deca-alanine can be predicted using a random pair of $F$ and $R$
work distributions contained in Fig. \ref{fig:CEver} (free energy
difference actually corresponds to the intersection point of the
$P_R(-W)$ and $P_F(W)$ distributions).  Such free energy differences
are roughly consistent with the free energy difference obtained from
the SMD work distributions (inset of Fig.  \ref{fig:CEver}) and those
reported in Ref. \onlinecite{procacci}.  A more detailed analysis
about free energy determination using the various FDA models is
provided in the Supplementary Material.

Given that the underlying decoupling ICA methods aim at separating
signals into non-Gaussian sources \cite{fastICA}, it is reassuring to
note that the $\mathcal{WAC}$ ICA approach is able to recover also
Gausian-like work distributions, such as those in the $F$ direction
(Fig. \ref{fig:CEver}). This is quantified by the skew and by the
difference between mean and median of the two $F$ work distributions
which are very small (skews: 0.0862 and -0.0142; difference between
mean and median: -0.0979 and 0.0249).  It should also be noted that
the ICA methods can give rise to convergence problems for small
samples sizes (this computational decoupling method usually requires
larger ``training sets" \cite{fastICA,stone_stICA}). We notice however
that in the present case good results are obtained for both
$\mathcal{W}$ ICA and $\mathcal{WAC}$ ICA bootstrapping even using
small sample sizes.  The positive performance of the small sample
$\mathcal{WAC}$ case is reassuring because this method can be used
even when the sample size in hand is not large enough for reliable use
of ICA methods.

\subsection{Physical Interpretation and collective variables
  orthogonal to the RC}
\label{sec:storytime}

Previous works have already made some connections between classical
statistical mechanics and nonequilibrium work relations
\cite{jarzynski06}. The time series studied here suggest that paths
exhibiting extreme work values, i.e., paths whose related work falls
in one of the tails of the measured work distribution, are associated
with stochastic dynamics which more strongly deviate from the
``average SPA model". These types of relationships were first
discovered by the data driven methods presented earlier (using only
the end-to-end distance RC).  Physics-based intuitive explanations
were then tested by analyzing the sequence of SMD all-atom
snapshots. The FDA methods revealed that there are clusters
(sub-populations) in our data sets.  The fact that the clusters appear
to have physical relevance (see discussion below) suggests that SPA
output may be useful for a functional classification
tool\cite{ferraty06} in systems where the underlying molecular details
are not well understood. We expand on this speculative idea below, but
first some established physical facts are presented.

When a process takes place far from equilibrium, classical statistical
mechanics predicts that paths on average dissipate more work than they
would if the process were carried out reversibly. We label these paths
as the ``uninteresting paths''.  We label those that microscopically
violate the second law of thermodynamics, i.e. with negative
dissipated work, as the ``interesting paths''\cite{jarzynski06}. The
labels are only used here to distinguish between the two cases in
terms of the contribution each of them gives to the exponential
average used for recovering free energy
differences\cite{jarzynski97}. We notice that in our system it is
difficult to distinguish between interesting and uninteresting paths
by visually inspecting the dynamics of the steered RC alone (see
bottom of Fig. \ref{fig:rcfail}). However there are other
configurational coordinates that we could use to make the differences
more apparent. Figure \ref{fig:oRC} shows some of these possible
configurational coordinates, namely the radius of gyration and the
number of hydrogen bonds in deca-alanine, as a function of the steered
RC. Specifically, in Fig. \ref{fig:oRC} we report the path population
mean $\mu$ and the curves that indicate the standard deviation, $\mu +
\sigma$ and $\mu - \sigma$, of these two quantities. Moreover the blue
lines denote a set of uninteresting paths, while the red lines denote
a set of interesting paths. Using the number of hydrogen bonds it is
difficult to distinguish between interesting and uninteresting paths,
but by using the radius of gyration, one observes that the interesting
paths form a tube that is systematically lower than the mean of the
entire population. It is noteworthy that the radius of gyration
associated, in turn, with interesting and uninteresting paths has a
significantly different behavior when $22 < Z < 26$ \Ang. When the
total amount of nonequilibrium work is used to index the radius of
gyration paths, a ``cluster-like" trend emerges. This indicates that
in this system the work produced during nonequilibrium processes can
help in classifying trajectories. Here, we aim to show that indexing
SPA models on the basis of the nonequilibrium work also induces
cluster-like behavior in the empirically measured SPA functions, and
that such a clustering presents a correlation with that revealed by
the radius of gyration. This observation is likely unique to
deca-alanine or to a small class of similar macromolecules. The
general concept, which we believe is applicable to a wider class of
systems, is that clusters of responses observed in the SPA model
correlate with some signature of a more complex (initially unobserved)
RC. In the case of simulations, one is free to investigate the
detailed trajectories to see if this is so and in wet-lab experiments
one may want to consider testing a variety of hypothesis to see if the
cluster of SPA models correlates to a physically interesting quantity.

In Fig. \ref{fig:ppDIFF} we show drift and diffusion coefficients of
two subsets of SPA models taken from a population of 60 estimated
models. The subsets correspond to SPA models whose work paths, or more
precisely the work paths corresponding to the underlying SMD paths
used to estimate the SPA model, fell into the two tails of the $R$
work distribution (note that in the figure the mean function
calculated over the 60 SPA model coefficient functions has been
subtracted from the individual functions).  Sampling the paths in the
tails of the work distribution allows us to select interesting
(smallest work values) and uninteresting paths (largest work
values). In the interesting paths, the drift function shows a faster
contraction at the \emph{beginning portion} of the path. A reasonable
explanation for this would be that, in the interesting paths, the
elongated ``random coil'' is aligned in such a way that it experiences
significantly more attractive forces from adjacent peptides (compared
to typical configuration consistent with the initial $Z$). The
converse applies to the uninteresting paths. Perhaps a more intriguing
trend is observed in the diffusion functions associated with
interesting paths. Initially they are low, but then increase rapidly
in a sort of ``transition'' that occurs around $Z \approx 26 \
\mathrm{\AA}$. It is interesting to note that the function describing
the noise magnitude undergoes this transition roughly at the same $Z$
value at which the radius of gyration paths ``diverge'' (compare
Figs. \ref{fig:oRC} and \ref{fig:ppDIFF}).  Some physicists might
attribute this fact to a meta-stable state that exists for $Z$ near
$26 \ \mathrm{\AA}$ \cite{ioanFR07}.

The possible connections existing between diffusion models and
structural features in nonequilibrium processes, as we have shown
here, suggests that incorporating dynamic response information (such
as that contained in this collection of SPA models) along with other
structural information may provide a useful tool in classifying
configurational trajectories where there are important unresolved (due
to lack of direct observation) configurational details that are
believed to be important \cite{zhuang02,xie04}. Identifying clusters
of dynamical responses could potentially assist in correlating SPA
model functions to both work and slowly evolving configurational
details using more sophisticated FDA techniques (e.g. functional
partial least squares or mixed linear models \cite{ferraty06}). This
is attractive in situations where small and large samples of
nonequilibrium paths are available. Physically, the clusters of curves
in Fig. \ref{fig:ppDIFF} suggest that, in this system, the microstates
associated with the slowly evolving degrees of freedom can be modeled
effectively as a continuum. For example, the radius of gyration
appears to take a continuous range of values for a given temperature
and fixed value of the RC. As a result of this, a continuous tube of
SPA models is likely observed because the (roughly) continuously
distributed radius of gyration reaction coordinate modulates the
dynamical response \footnote{Of course, many other factors are
relevant in determining the dynamical response but the basic idea
remains the same.}.  In more complex macromolecules, one should be
warned that the SPA models estimated will likely depend on the
configurational degrees of freedom in a more complicated fashion, but
if they do measurably modulate the dynamics (e.g. see
Ref. \onlinecite{voth07}), then our technique may be helpful (but the
nonequilirium work will probably not be approximated as accurately as
it was here).

\section{Conclusions and perspectives}
\label{sec:conclusions}

We have presented a surrogate process approximation (SPA)
method\cite{SPA1} based on a diffusion process to model nonequilibrium
paths of a given reaction coordinate. We have applied the methodology
to the nonequilibrium dynamics of the end-to-end distance of a
deca-alanine peptide at finite temperature produced by steered
molecular dynamics (SMD) simulations. Special attention has been
devoted to the refolding process of deca-alanine, that was previously
demonstrated to yield markedly non-Gaussian work
distributions\cite{procacci}. For modeling the dynamical behavior of
the reaction coordinate of deca-alanine, two strategies have been
proposed. One is based on the modleling of one SMD path through a
series of local diffusion models, each describing a limited interval
of the path. These local models are then stitched to give the global
diffusion model (SPA model) for the given SMD path. A collection of
global SPA models, modeled on various statistically independent SMD
paths, have been then used to increase the statistical sampling, that
is to produce a large amount of surrogate paths. Unluckily, these
surrogate paths have been demonstrated to furnish excessively broad
work distributions. In order to correlate the dynamics of the reaction
coordinate resulting from the SPA models with the work paths observed
in the underlying SMD paths, we have turned to functional data
analysis\cite{ramsay}. This procedure has allowed us to recover good
quality work distributions considering the relatively small number of
SMD paths used for the modeling. This is indeed one of the goals of
our study: when the number of direct observations (from experiments or
calculations) is small, our approach allows one to obtain reliable
estimation of the work distribution shape. Given the fact that several
recent methods for estimating free energy
differences\cite{jarzynski97,crooks98} depend on approximations of the
work distributions obtained from a finite collection of nonequilibrium
realizations along a given reaction coordinate, any method for
improving the quality of the work distributions may provide
considerable help. Moreover, the determination of the work
distribution shape can also furnish significant insights on the
underlying dynamics of nonequilibrium processes. Taking advantage of
this possibility, we are currently investigating strategies for
supplying work distribution reweighting algorithms (see, e.g., Eq. 6
of Ref.  \onlinecite{chelliarxiv07}) with SPA protocols.

A further outcome of this study is to reveal the possibility of using
SPA models to detect stochastic dynamical features of a system,
e.g. trends in the drift or diffusion coefficient of an approximating
SDE, from noisy reaction coordinate time series (obtained from
nonequilibrium simulations).  In fact, the dynamical behavior of
slowly evolving degrees of freedom (or of collective variables
orthogonal to the chosen reaction coordinate), has been shown to be
emphasized by the SPA models. Evidence of this can be found in the
diverse ``population" of drift and diffusion functions associated with
the collection of estimated SPA models (see Figs. \ref{fig:multmodel},
\ref{fig:oRC} and \ref{fig:ppDIFF}). It should be carefully noted that
the drift and diffusion functions of a single SPA model alone
(estimated using a single SMD path) can hardly reveal the occurrence
of peculiar heterogeneities in the dynamics of the system. On the
other hand, such features can be detected by analyzing clusters in a
collection of (estimated) SPA drift and diffusion coefficient
functions.  At least this is the case observed in deca-alanine, where
the dynamics of the radius of gyration { appears to be significantly
correlated to both drift and diffusion functions.  The radius of
gyration was shown to readily identify work paths which transiently
violated the second law of thermodynamics.  It should be noted that,
in a real experiment, a careful analysis of additional degrees of
freedom, as we have done for deca-alanine, may be difficult or simply
not feasible. The methods presented here provide an indirect way of
quantifying the effect of difficult to measure degrees of freedom
utilizing dynamical information contained in a collection of time
series of a simple reaction coordinate.

In summary we have demonstrated that the measured diffusion functions
offer an alternative method for using ``noise" (estimated on a
pathwise basis) in nonequilibrium simulations/experiments to detect
subtle small scale transitions.  One possible future application of
this observation would be to develop a sampling method which
constructs a basis (e.g.  functional PCA) from a few ``entire"
pullings (i.e. span the entire RC range of interest). If after this
step is completed, transitions regions become apparent (or the
location of transitions is known from established physical theories),
one could exploit this finding and run a larger batch of ``short" SMD
simulations/experiments. By this we mean that the SMD
simulations/experiments would only need to be carried out until just
after the transition region of the RC. In this way, one could project
onto the previously determined (empirical) basis and use the larger
set of new trajectories to get a refined estimate of the frequency of
certain important small scale molecular events by analyzing the
distribution of the projections onto this basis.

\begin{acknowledgments}
Acknowledgements The work of C.P.C. was supported by an NSF grants
\#DMS-0240058 and \#ACI-0325081 and R.C.  by the European Union, grant
\# RII3-CT-2003-506350.
\end{acknowledgments}


\newpage

\begin{table*} [ht]
\caption{\label{tab:Wstats}Work Statistics. The case labelled ``SMD"
corresponds to using SMD trajectories to evaluate work at the terminal
endpoint. In all of the FDA bootstrapping cases, $10^4$ bootstrapped
processes have been used. The number in parenthesis indicates the
number of SMD paths used to calibrate each collection of SPA models
(in the SMD cases this corresponds to the number of paths where work
was measured directly from).  In this Table only, the larger sample
cases build incrementally from the smaller sample cases, e.g.
$\mathcal{W}$ (30) contains the 15 paths from $\mathcal{W}$ (15).  }
\begin{tabular}{lcccccc}
 Case \ci $\hat{\mu}$ \ci $\hat{\mu}$-Median \ci Mode \ci  $\hat{\sigma}$ \ci Skew \ci
 Kurtosis \\ \hline
SMD (15) \ci -77.785  \ci  2.478  \ci -75.270  \ci 13.390  \ci
1.491 \ci 4.915  \\ \hline

SMD (30) \ci -79.362  \ci  1.915  \ci -82.014  \ci 11.827  \ci
1.210 \ci 5.254  \\ \hline

SMD (60) \ci -77.481  \ci  1.424  \ci -82.795  \ci 12.166  \ci
1.235 \ci 5.382  \\ \hline

SMD (100) \ci -78.123  \ci  2.642  \ci -87.845  \ci 12.154  \ci
0.993  \ci 4.583  \\ \hline

SMD (1000) \ci -77.661  \ci  1.301  \ci -84.889  \ci 11.844  \ci
1.127  \ci 5.385  \\ \hline

$\mathcal{WAC}$ ICA (15) \ci -80.591  \ci  2.677  \ci -88.129  \ci
13.408 \ci 0.856 \ci 3.228  \\ \hline

$\mathcal{WAC}$ ICA (30) \ci -80.100  \ci  0.789  \ci -82.127  \ci
11.634  \ci  0.358  \ci 2.994  \\ \hline

$\mathcal{WAC}$ ICA (60) \ci -80.151  \ci  1.366  \ci -86.195  \ci
12.165  \ci  0.510  \ci 2.985  \\ \hline

$\mathcal{W}$ ICA (15) \ci -80.736  \ci  2.105  \ci -85.370  \ci
13.428  \ci  1.032  \ci 4.233  \\ \hline

$\mathcal{W}$ ICA (30) \ci -81.688  \ci  1.300  \ci -84.595  \ci
11.538  \ci  0.948  \ci 4.900  \\ \hline

$\mathcal{W}$ ICA (60) \ci -80.042  \ci  1.119  \ci -82.647  \ci
12.161  \ci  0.736  \ci 4.279  \\ \hline

$\mathcal{WAC}$ (15) \ci -79.560  \ci  3.027  \ci -83.593  \ci
12.693  \ci  1.024  \ci 3.726  \\ \hline

$\mathcal{WAC}$ (30)\ci -78.427  \ci  0.563  \ci -80.560  \ci
11.351 \ci  0.110  \ci 3.417  \\ \hline

$\mathcal{WAC}$ (60)\ci -75.564  \ci  0.980  \ci -82.097  \ci
11.561 \ci  0.306  \ci 2.859  \\ \hline

$\mathcal{W}$ (15)\ci -73.196  \ci  2.376  \ci -86.172  \ci 14.201
\ci  0.240  \ci 1.832  \\ \hline

$\mathcal{W}$ (30)\ci -76.847  \ci  0.426  \ci -74.422  \ci 10.621
\ci  0.338  \ci 3.488  \\ \hline

$\mathcal{W}$ (60)\ci -75.353  \ci  1.416  \ci -83.797  \ci 12.163
\ci  0.623  \ci 3.305  \\ \hline

\end{tabular}
\end{table*}

\clearpage
\newpage

CAPTIONS TO THE FIGURES

\begin{figure} [ht]  
\caption{ \label{fig:rcfail} (Color online) Snapshots of deca-alanine
during two independent refolding processes obtained from steered
molecular dynamics simulations published in Ref.
\onlinecite{procacci}. In the right side we report snapshots from a
process where misfolding occurs and the work done on the system is
greater than that obtained from a reversible process. In the left side
we report snapshots from a process that instead microscopically
violates the second law of thermodynamics \cite{jarzynski06}. The
center panel contains the reaction coordinate $Z(\tau)$ for both
simulations as a function of time.  In the bottom panel we plot
$Z(\tau)-\lambda (\tau)$ , where $\lambda(\tau)$ is the deterministic
protocol used for steering the reaction coordinate. The snapshots were
generated with the VMD program \cite{vmdref}.}
\end{figure}

\begin{figure} [ht]  
\caption{ \label{fig:multmodel} (Color online) Drift and diffusion
functions (panels A and B, respectively) estimated using 15 SMD paths
in the $F$ and $R$ directions (black and red curves,
respectively). The sampling step of the SMD paths is 150 fs. }
\end{figure}

\begin{figure} [ht]  
\caption{ \label{fig:worktraj} (Color online) Mean $\mu$ and standard
deviation $\sigma$ for $R$ work paths using the SMD and SPA
processes. The {curves labelled with} $W_i$ correspond to a few
randomly {chosen} work {paths (see} legend).}
\end{figure}

\begin{figure} [ht]  
\caption{ \label{fig:Rdens} (Color online) Work distributions in the
$R$ direction obtained from 1000 SMD simulations and from the FDA
bootstrapping methods (see legend).  The circles indicate the work
values corresponding to the 15 SMD paths used for calibrating the SPA
models used in FDA bootstrapping.}
\end{figure}

\begin{figure} [ht]  
\caption{ \label{fig:CEver} (Color online) $\mathcal{WAC}$ ICA work
distributions in the $R$ and $F$ directions (3 solid curves and 2
dashed curves, respectively). In the inset the $R$ and $F$ work
distributions obtained from the SMD simulations are reported { (black
and green histograms, respectively).} The work value at the
intersection point of these last work distributions corresponds to the
free energy difference (91.7 kJ mol$^{-1}$).  We stress that each FDA
bootstrapping (approximate) work density used different iid SMD paths
as the underlying data. For example in the $R$ direction 45 total SPA
models were estimated and this collection was split into 3 sets of
15. These batches were used to create the approximate $R$ work
densities shown here. }
\end{figure}

\begin{figure} [ht]  
\caption{ \label{fig:oRC} (Color online) Alternative RCs, i.e.  radius
of gyration (panel A) and number of hydrogen bonds in deca-alanine
(panel B), as a function of the steered RC (from $R$ SMD
simulations). The black dashed line denotes the mean population curve
$\mu$ calculated using 100 SMD paths.  The solid black lines denote
the standard deviation curves $\mu + \sigma$ and $\mu - \sigma$. 10
``interesting" and 50 ``uninteresting" paths are plotted in red and
blue, respectively.  Both quantities were calculated using the VMD
program \cite{vmdref}.  The radius of gyration was computed using all
atoms of deca-alanine. The hydrogen bonds were computed using the VMD
subroutine $\mathrm{Hbonds}$ using a 3 \Ang HN$\cdots$OC bond distance
and $20^o$ O$\cdots$N-H angle cutoffs.  }
\end{figure}

\begin{figure} [ht]  
\caption{ \label{fig:ppDIFF} (Color online) Drift (panel A) and
diffusion (panel B) functions (with population mean subtracted) of SPA
models calibrated on interesting and uninteresting $R$ SMD paths
(solid and dashed lines, respectively). }
\end{figure}

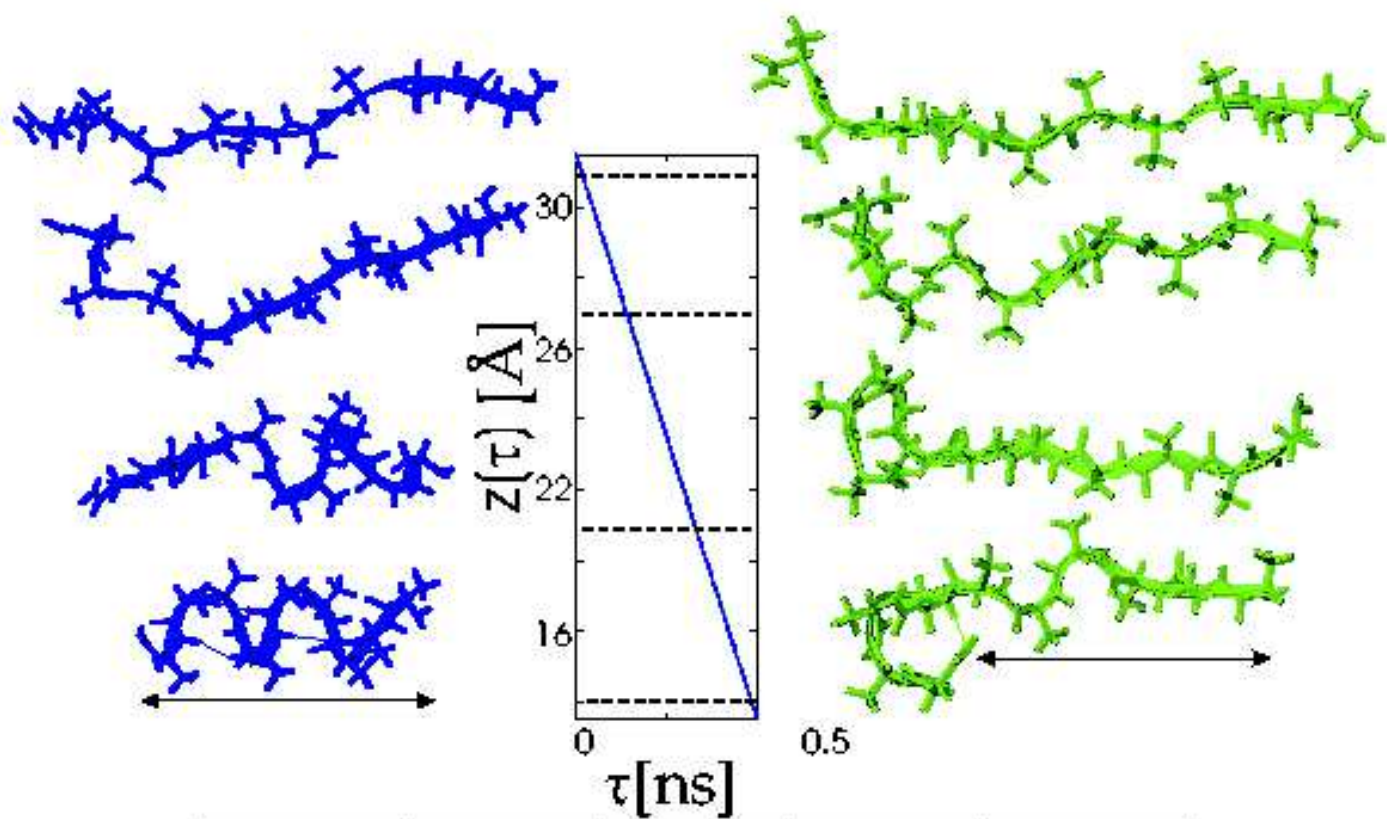
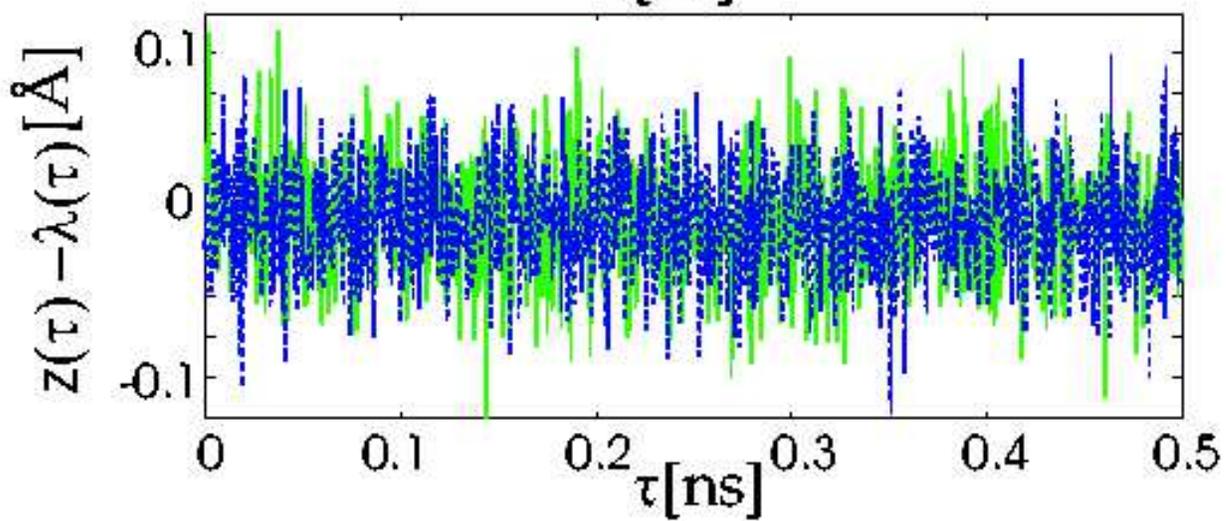
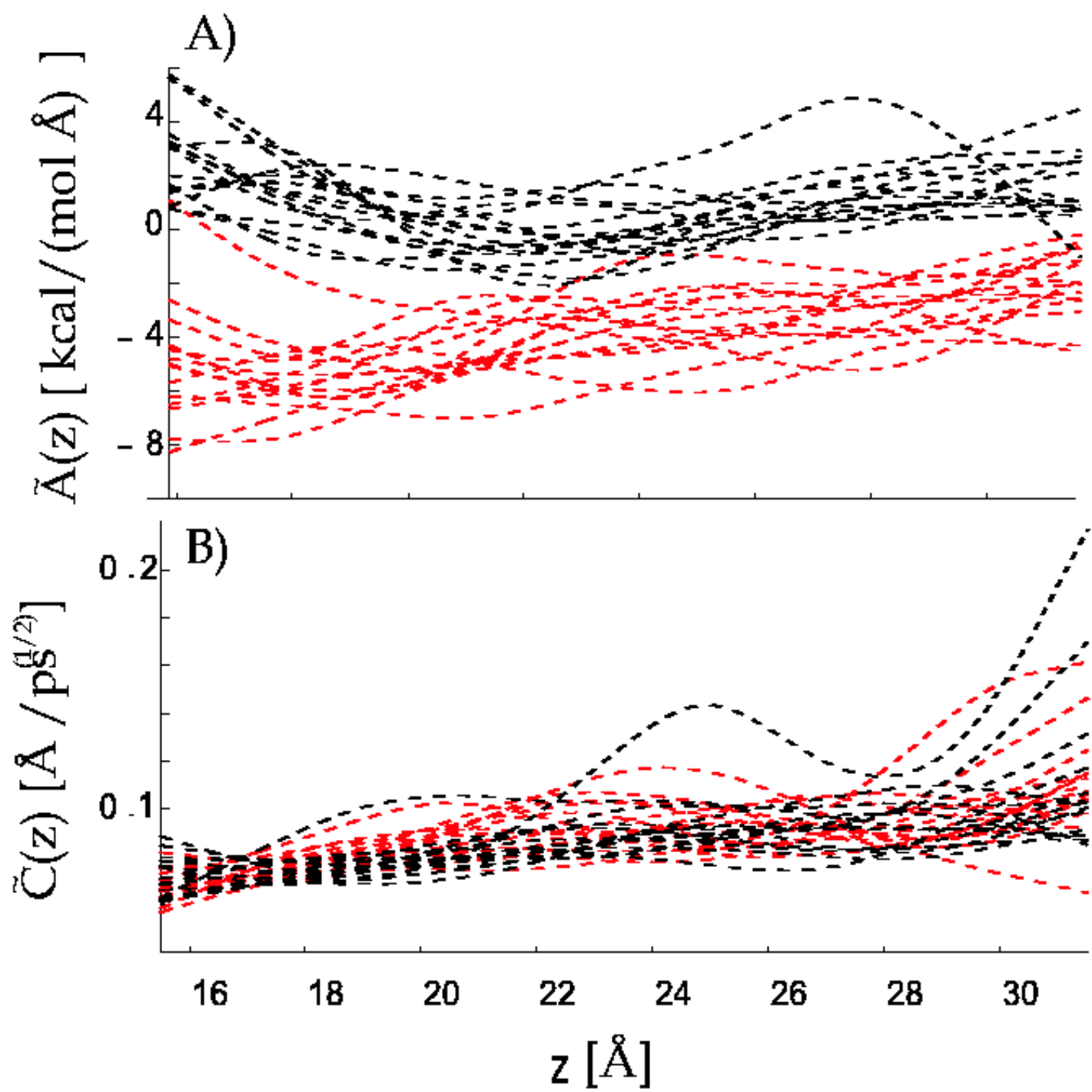
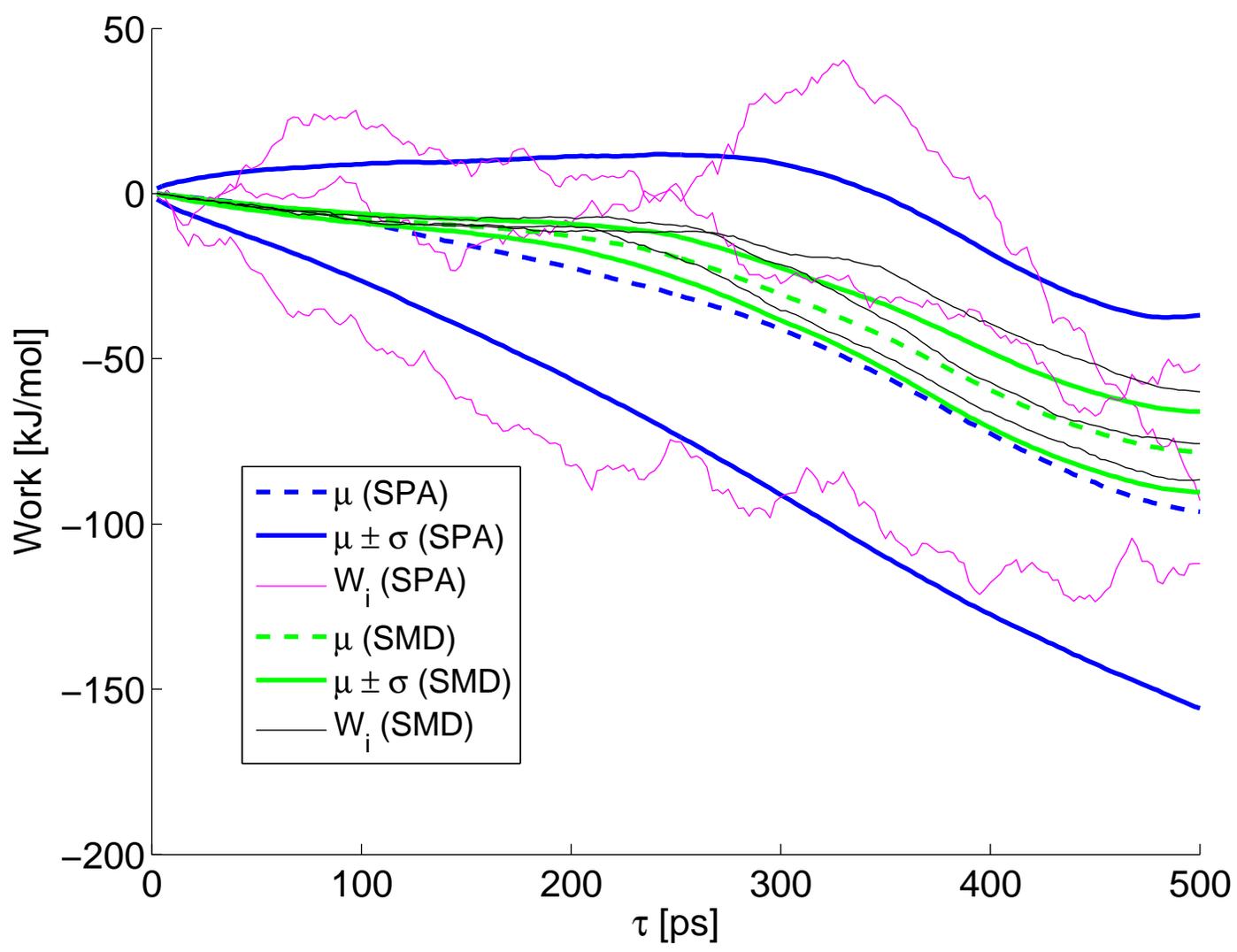
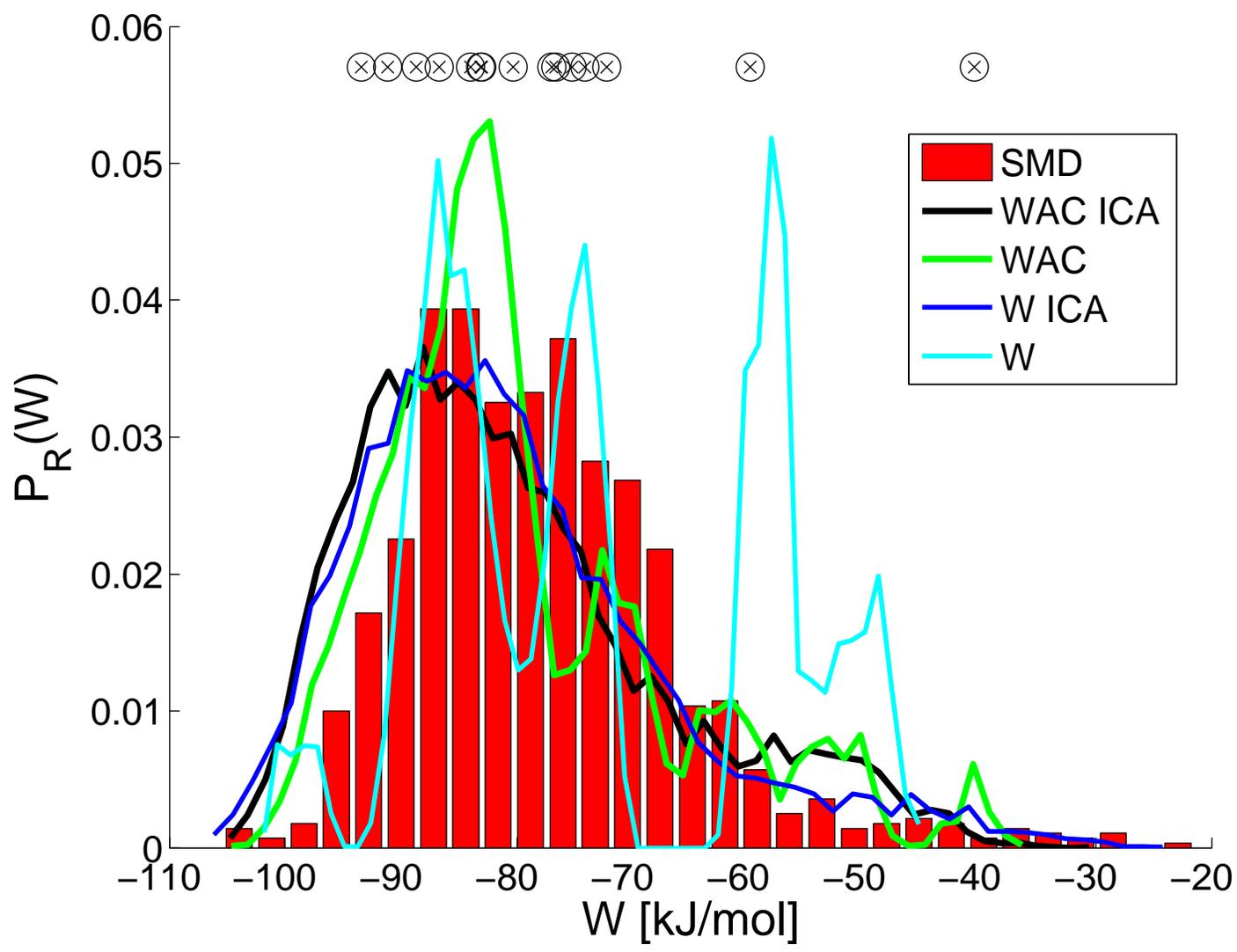
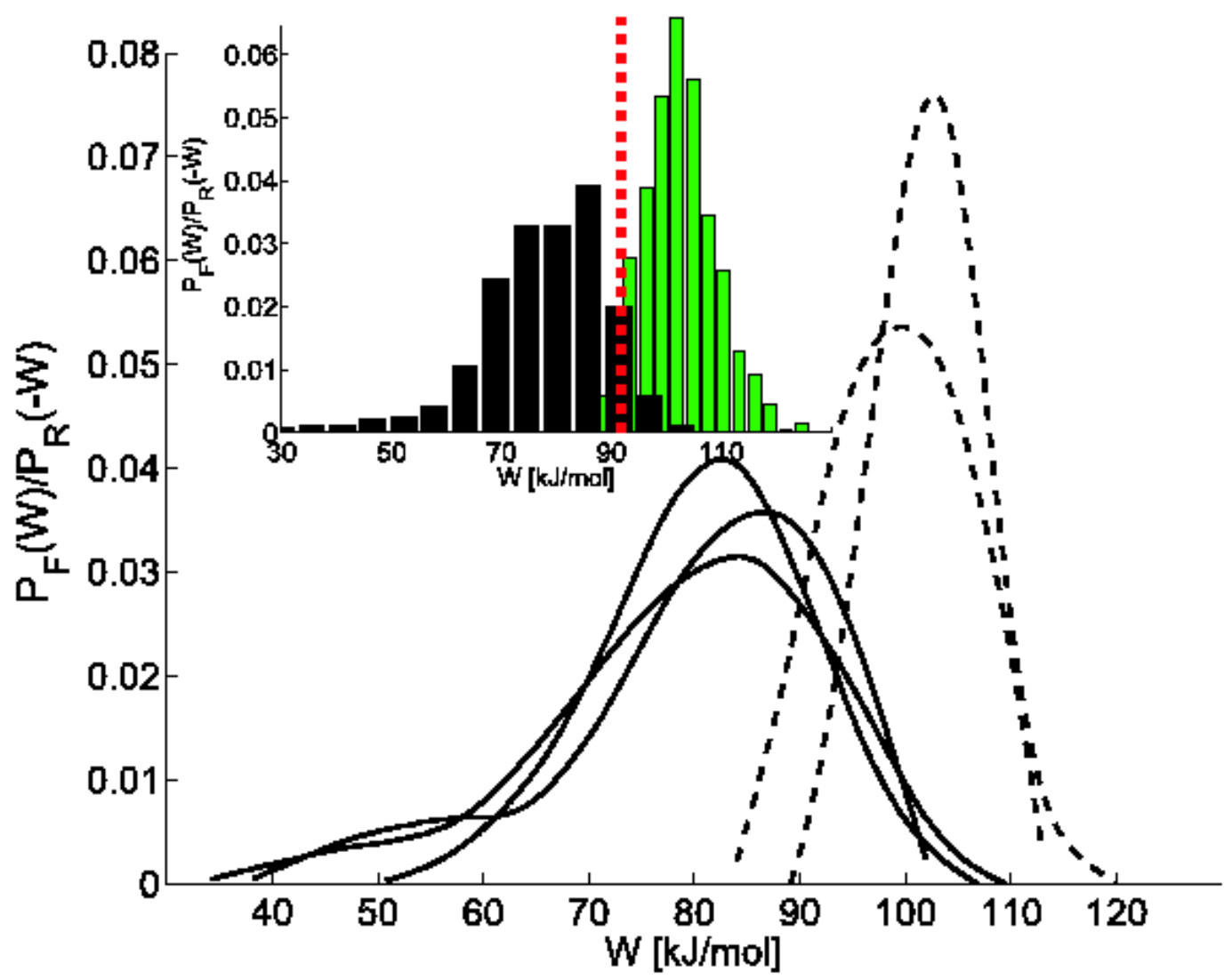
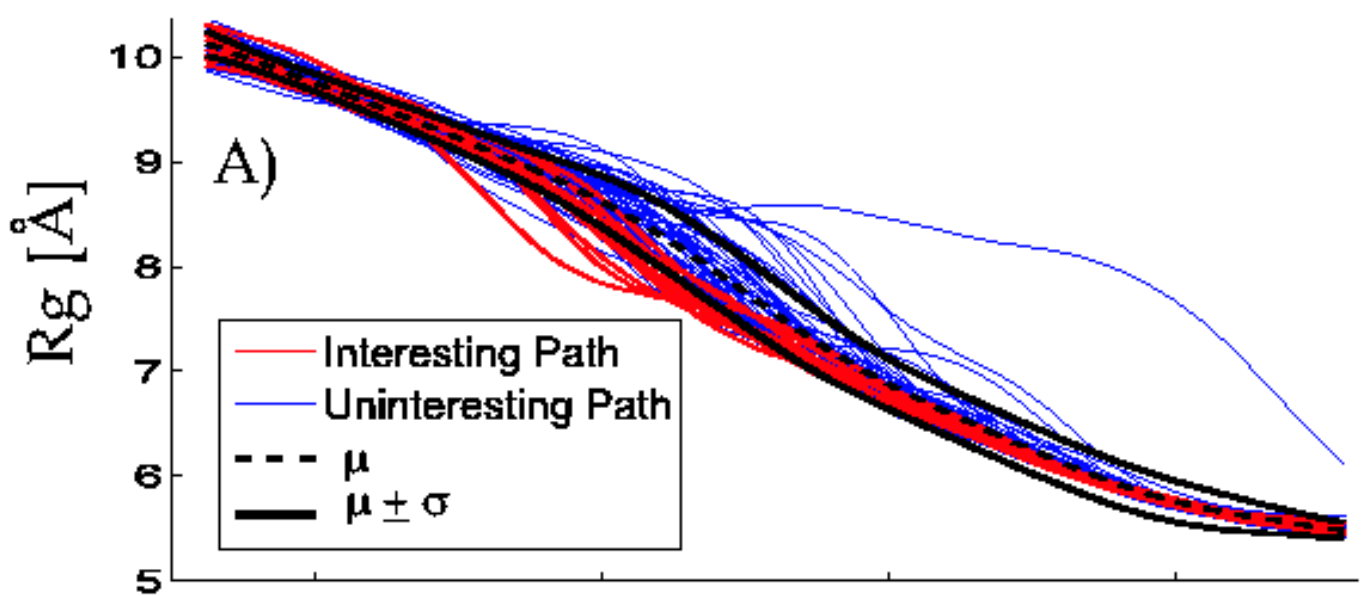
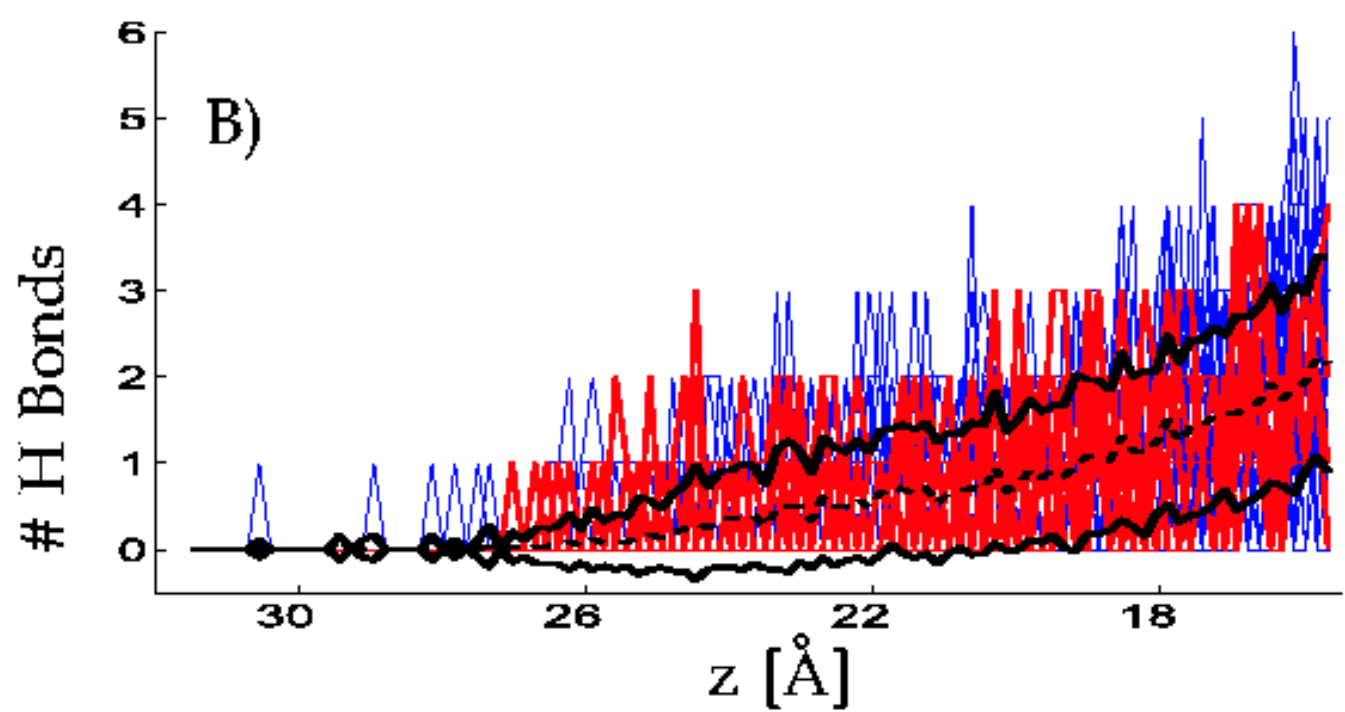
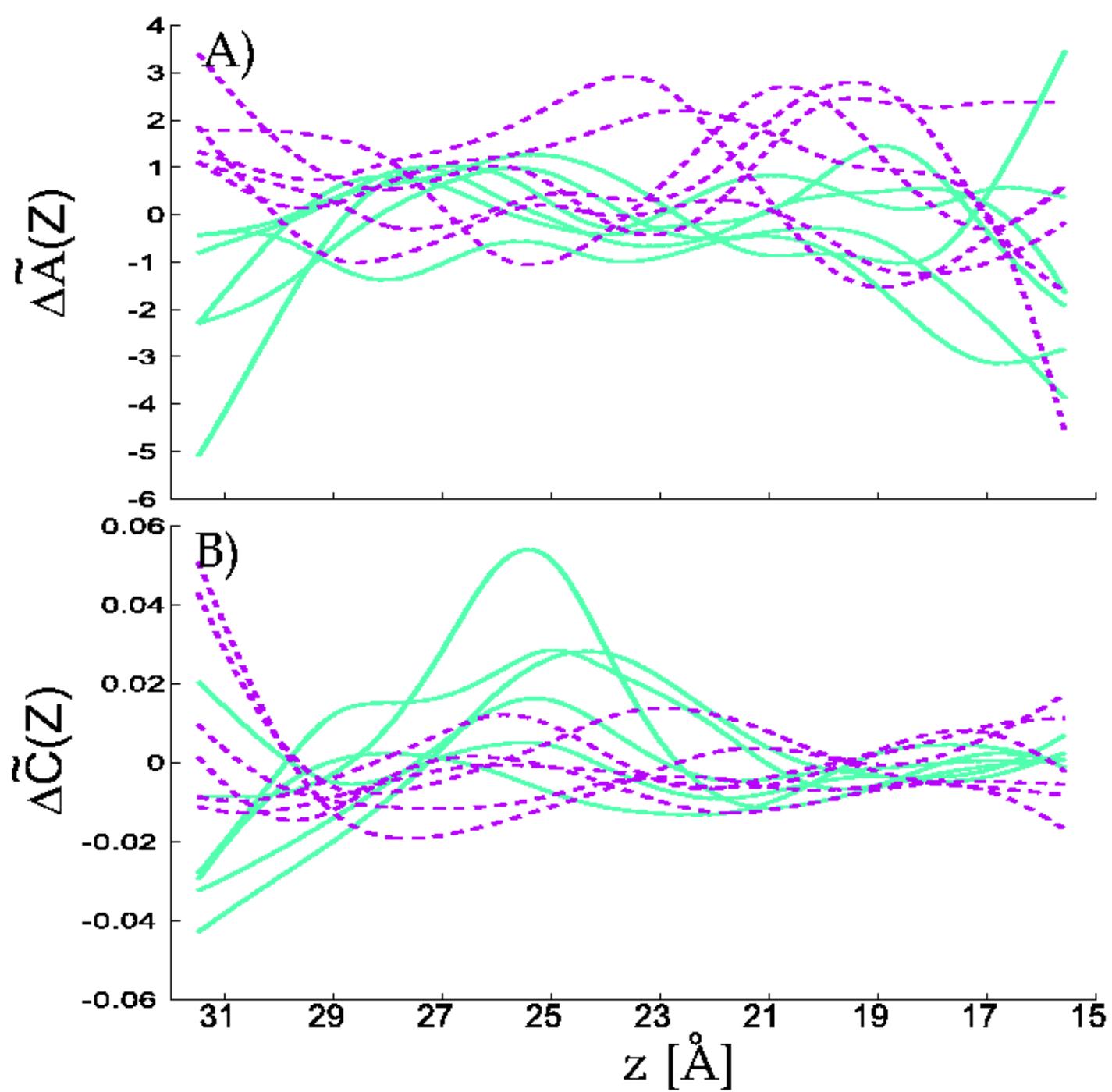
\end{document}